\documentclass[12pt]{article}
\usepackage[margin=1in]{geometry}
\usepackage{amsmath, amssymb, bm}
\usepackage{graphicx}
\usepackage{physics}
\usepackage{float}
\usepackage{cite}
\usepackage{setspace}
\usepackage{tikz}
\usetikzlibrary{patterns}
\usepackage{xcolor}
\title{ Spatial Correlations Restore Zwanzig’s Mean-Field Diffusion Result in Rugged Energy Landscapes
}
\author{\textbf{Biman Bagchi}\\
\small Solid State and Structural Chemistry Unit,\\
\small Indian Institute of Science, Bengaluru 560012, India}

\date {}
\begin{document}
\maketitle
\begin{abstract}
Transport in disordered environments is often controlled not by typical fluctuations but by rare, extreme events that dominate long-time dynamics. In such settings, Zwanzig’s classic mean-field theory predicts that energetic roughness reduces the diffusion coefficient by an exponential factor governed solely by the variance of the disorder. However, this prediction breaks down in uncorrelated Gaussian landscapes, where rare but deep multi-site traps dominate transport and lead to a much stronger suppression of diffusion. Here, we present a unified theoretical framework that clarifies both the origin of this breakdown and its resolution. We show that Zwanzig’s local averaging can be interpreted as a Gaussian cumulant expansion whose validity is destroyed by uncorrelated disorder through the emergence of extreme trapping events. Introducing Gaussian spatial correlations fundamentally reshapes the landscape: roughness increments become smoother, asymmetric multi-site traps are suppressed, and the statistics of escape pathways are regularized. As a result, Zwanzig’s exponential scaling is recovered.
We provide an explicit analytical derivation demonstrating how spatial correlations modify trap statistics and restore mean-field diffusion, complemented by illustrative numerical examples showing the dramatic reduction of escape times in correlated landscapes.
\end{abstract}
%

\section{Introduction}

Diffusion in a rugged free-energy landscape is a recurring theme in chemical physics, biophysics, and the physics of disordered materials. 
The concept arises naturally whenever a particle or molecular entity moves through an environment whose local energetic environment fluctuates on nanoscopic length scales. 
Examples span an unusually wide range of systems: ion transport in glasses, polymer segmental dynamics, conformational diffusion in proteins, catalysis and enzyme motions, and the sliding or hopping motion of proteins along DNA. 
In each of these contexts, the effective potential experienced by the diffusing entity contains numerous small barriers and wells generated by microscopic structural heterogeneities, and the statistics of these fluctuations strongly affect long-time transport.

The modern theoretical treatment of diffusion in such landscapes began with the work of Zwanzig,\cite{zwanzig_pnas_1988} 
who analyzed motion in a one-dimensional potential composed of a smooth background plus a random ``roughness'' term. 
Zwanzig obtained the strikingly simple prediction that the ruggedness renormalizes the diffusion constant by an exponential factor depending only on the variance of the disorder.  
This observation has been widely used in diverse fields because it suggests that the complicated details of the potential can be replaced with a single coarse parameter.

However, subsequent analytical and simulation studies  [2-5],
demonstrated that Zwanzig’s expression has limited validity.   
When the roughness values are taken to be completely uncorrelated from site to site, the diffusion constant can be reduced by several orders of magnitude relative to Zwanzig’s prediction.  
The essential physical reason for this discrepancy is the emergence of rare but extremely deep multi-site traps, particularly the so-called three-site traps (TSTs), which dominate the long-time transport in an uncorrelated landscape.  
These pathological configurations are effectively eliminated in Zwanzig’s calculation because his derivation relies on a local smoothing or coarse-graining step that implicitly presumes correlated disorder.

The relevance of rugged landscape diffusion is especially clear in biomolecular systems.  
The motion of restriction enzymes and transcription factors along DNA is directly influenced by a sequence-dependent binding landscape.  
Single-molecule experiments by Blainey, van Oijen, Banerjee, Bagchi, and Xie\cite{blainey_xie_nsmb}
demonstrated that proteins execute sliding, hopping, and switching between one- and three-dimensional modes of motion in a free-energy landscape whose ruggedness reflects base-pair sequence heterogeneity.
Similarly, conformational transitions in proteins and enzymes occur on landscape topographies containing many small-amplitude barriers arising from side-chain packing and solvent-induced fluctuations.  
Rugged landscapes also appear in polymer dynamics, glassy relaxation, and in the energy-landscape description of supercooled liquids developed by Stillinger and Weber,\cite{stillinger_weber}
extended by Heuer,\cite{heuer_review}
and incorporated into theoretical frameworks such as the random first-order transition approach.\cite{wolynes_rfpot}
Across these diverse problems, understanding how microscopic roughness controls macroscopic transport remains a central theme.

A key insight emerging from the work of Banerjee, Biswas, Seki, and Bagchi (BBSB) is that the validity of Zwanzig’s mean-field result depends sensitively on the \emph{spatial structure} of the disorder.  
To quantify this structure, they introduced Gaussian spatial correlations into the roughness, a form long used in astrophysics to describe turbulent density fields and the morphology of interstellar media.\cite{gaussian_astrophysics1, gaussian_astrophysics2}
Remarkably, even modest correlations dramatically suppress the occurrence of TSTs and restore Zwanzig’s exponential diffusion law.  
Thus, the apparent breakdown of the mean-field picture is not intrinsic to rugged landscapes themselves, but rather to the unrealistic assumption of uncorrelated disorder.  
Real physical systems—DNA sequences, polymer chains, hydrogen-bond networks, and disordered solids—possess finite correlation lengths arising from molecular connectivity and intermolecular forces, making the correlated model far more realistic.
In the numerical work of Banerjee {\it et al.} (BBSB), the rugged landscape was 
constructed by drawing site energies from a Gaussian distribution with variance 
$\epsilon^2$, a choice that is natural for a high-dimensional disordered environment 
where the local energy results from the sum of many weak contributions.  
This Gaussian distribution in \emph{energy space} is distinct from the original 
example given by Zwanzig, who illustrated his ideas using a smooth but artificial 
cosine–based potential containing many harmonics.  
Zwanzig’s construction was never intended as a literal physical model of disorder; 
it simply allowed him to carry out the local averaging explicitly.  
In contrast, the BSB approach begins with a physically motivated assumption: 
the disorder itself is Gaussian-distributed from the outset, consistent with a 
central-limit–type argument for landscape heterogeneity.

It is important to emphasize that Zwanzig’s “local averaging” procedure is 
mathematically equivalent to the assumption that the fluctuations of the roughness increments 
are Gaussian and sufficiently small that the first nontrivial cumulant—the variance—
dominates.  
Thus, although his illustrative $U(x)$ was expressed in terms of cosine functions, 
the underlying statistical assumption is Gaussian: the coarse-grained increments 
are treated as Gaussian random variables.  
This leads to an interesting structural parallel between the two approaches: 
BSB impose Gaussian statistics at the level of site energies, while Zwanzig’s 
coarse-graining effectively imposes Gaussian statistics on the \emph{increments} 
of his roughness function.  
In principle, one may even contemplate a unified framework in which the Gaussian 
energy distribution (over sites) and the Gaussian spatial correlations 
(over real-space distances) are coupled, although such an interdependence was not 
explored in the original studies.  
The present work clarifies how these different Gaussian assumptions relate to one 
another and how spatial correlations in real space determine whether the cumulant 
expansion implicit in Zwanzig’s analysis remains valid.

The goal of this manuscript is to assemble a coherent and self-contained account of how spatial 
correlations modify rugged landscape diffusion, beginning from Zwanzig’s original formulation, moving 
through the breakdown identified by Banerjee, Biswas, Seki, and Bagchi, and culminating in the demonstration that Gaussian correlations restore mean-field behavior.  
We also provide a clear derivation of how correlations alter the statistics of roughness increments, reduce multi-site trapping, and modify the long-time diffusion constant.  
This unified perspective clarifies the true physical origin of the Zwanzig scaling and highlights the broader importance of correlated disorder in chemical and biological transport processes.

The present manuscript develops a framework for understanding how spatial correlations in the roughness restore Zwanzig’s mean-field expression.


\section{Zwanzig's MFPT route and the origin of the exponential renormalization}
\label{subsec:Zwanzig_MFPT_clean}

We briefly recall Zwanzig's derivation of the effective diffusion constant for
overdamped motion in a one-dimensional rough potential, emphasizing (i) the
\emph{exact} mean-first-passage-time (MFPT) identity, (ii) the single
\emph{approximation} used to obtain a closed form, and (iii) the precise role
played by Gaussian averaging.

Consider overdamped diffusion in a static potential
\begin{equation}
U(x)=U_0(x)+\eta(x),
\end{equation}
with bare diffusivity $D_0$ (in the absence of $\eta$).  The exact MFPT formula
for diffusion in one dimension implies an exact expression for the long-time
effective diffusivity (Zwanzig's Eq.~(5) in Ref.~\cite{Zwanzig1988}):
\begin{equation}
D_{\mathrm{eff}}
=
\frac{D_0}{
\displaystyle
\lim_{L\to\infty}
\frac{1}{L^2}\int_{0}^{L}\!dx\, e^{\beta U(x)}
\int_{0}^{L}\!dy\, e^{-\beta U(y)}
}.
\label{eq:Zw_MFPT_exact}
\end{equation}
Equation~\eqref{eq:Zw_MFPT_exact} is \emph{exact} for one-dimensional diffusion
in a static potential; no approximation has yet been made.

\paragraph{Key approximation: spatial self-averaging (factorization).}
When the random component $\eta(x)$ has a short spatial correlation length
(compared to $L$), the two integrals in Eq.~\eqref{eq:Zw_MFPT_exact} become
self-averaging in the limit $L\to\infty$.
Operationally, one replaces the spatial averages by ensemble averages:
\begin{equation}
\frac{1}{L}\int_{0}^{L}\!dx\, e^{\pm \beta \eta(x)}
\;\longrightarrow\;
\left\langle e^{\pm \beta \eta}\right\rangle,
\qquad (L\to\infty),
\label{eq:self_averaging}
\end{equation}
and, crucially, assumes that the product of the two large-$L$ averages
factorizes:
\begin{equation}
\left(\frac{1}{L}\int_{0}^{L}\!dx\, e^{\beta \eta(x)}\right)
\left(\frac{1}{L}\int_{0}^{L}\!dy\, e^{-\beta \eta(y)}\right)
\;\approx\;
\left\langle e^{\beta \eta}\right\rangle
\left\langle e^{-\beta \eta}\right\rangle.
\label{eq:factorization_step}
\end{equation}
This is the \emph{mean-field-like} step: it neglects residual correlations
between the two spatial integrals that originate from the joint statistics of
$\eta(x)$ and $\eta(y)$ in the MFPT expression.

With $U_0(x)$ taken to be slowly varying (or constant on the roughness scale),
Eqs.~\eqref{eq:self_averaging}--\eqref{eq:factorization_step} reduce
Eq.~\eqref{eq:Zw_MFPT_exact} to
\begin{equation}
D_{\mathrm{eff}}
\approx
\frac{D_0}{
\left\langle e^{\beta \eta}\right\rangle
\left\langle e^{-\beta \eta}\right\rangle
}.
\label{eq:Zw_reduced}
\end{equation}

Assume the roughness is Gaussian with zero mean and variance
$\langle \eta^2\rangle=\epsilon^2$.
Then the Gaussian moment identity gives
\begin{equation}
\left\langle e^{\alpha \eta}\right\rangle
=
\exp\!\left(\frac{\alpha^2\langle \eta^2\rangle}{2}\right)
=
\exp\!\left(\frac{\alpha^2\epsilon^2}{2}\right),
\label{eq:Gaussian_moment}
\end{equation}
which is exact for Gaussian $\eta$.  The factor $1/2$ enters \emph{here}, through
Eq.~\eqref{eq:Gaussian_moment}.  Applying Eq.~\eqref{eq:Gaussian_moment} to
Eq.~\eqref{eq:Zw_reduced} yields
\begin{equation}
\left\langle e^{\beta \eta}\right\rangle
=
\exp\!\left(\frac{\beta^2\epsilon^2}{2}\right),
\qquad
\left\langle e^{-\beta \eta}\right\rangle
=
\exp\!\left(\frac{\beta^2\epsilon^2}{2}\right),
\end{equation}
so that their product is $\exp(\beta^2\epsilon^2)$, and therefore
\begin{equation}
D_{\mathrm{eff}}
\approx
D_0\,\exp\!\left[-\beta^2\epsilon^2\right].
\label{eq:Zw_final}
\end{equation}
Thus, while the Gaussian identity contains a factor $1/2$, the final exponent
in Eq.~\eqref{eq:Zw_final} contains no $1/2$ because it arises from the
\emph{product} of two Gaussian averages in Eq.~\eqref{eq:Zw_reduced}.

Equation~\eqref{eq:Zw_final} follows from (i) the exact MFPT identity
Eq.~\eqref{eq:Zw_MFPT_exact} and (ii) the factorization/self-averaging step
Eq.~\eqref{eq:factorization_step}, which becomes accurate when the roughness is
short-range correlated and the observation length $L$ is large.
The Gaussian moment evaluation in Eq.~\eqref{eq:Gaussian_moment} is exact and is
\emph{not} an additional approximation.
In later sections we will discuss how discrete-lattice models and rare-event
multi-site traps introduce corrections beyond Eq.~\eqref{eq:factorization_step}
when transport occurs on a lattice with uncorrelated site-to-site energetics.

\medskip
\noindent
\textbf{Relation between continuous and discrete descriptions.}
Before proceeding, it is useful to clarify the relationship between
continuous and discrete formulations of diffusion in rugged landscapes.
A discrete random walk with sufficiently smooth spatial variation of
transition rates admits a well-defined continuum limit, while a
continuous potential may always be discretized on a lattice for
numerical or analytical convenience.
In this sense, the two descriptions are not fundamentally distinct;
rather, they represent different coarse-grained realizations of the
same underlying stochastic dynamics.

\medskip
\noindent
The effective diffusivity derived in Sec.~II assumes that the roughness
$\eta(x)$ is a spatially continuous random field with finite correlation
length and that the large-scale transport can be described through the
self-averaging/factorization step of Eq.~(\ref{eq:factorization_step}).
When a discrete lattice model is constructed so that its continuum limit
reproduces these assumptions (e.g., slowly varying energetics and finite
correlation length), one indeed recovers Zwanzig’s exponential scaling.

\medskip
\noindent
However, when the lattice model contains site-to-site energetics that
are uncorrelated or vary abruptly on the scale of a single lattice
spacing, the statistics of local escape times can become dominated by
rare extreme configurations.
In one dimension, where traversal times add sequentially, such rare
configurations may exert a disproportionate influence on the harmonic
mean that governs long-time transport.
The resulting deviations from Eq.~(\ref{eq:Zw_final}) therefore reflect
not a contradiction of the continuous theory, but the emergence of
additional rare-event physics that becomes explicit in discrete
representations with short-range disorder.
In the next section we formulate this lattice problem explicitly and
derive the corresponding exact expression for the diffusion constant.
%
\section{Discrete Random Landscapes and Rare-Event Contributions}

In order to understand deviations from Eq.~(\ref{eq:Zw_final})
in certain numerical simulations, we now consider a discrete
representation of diffusion on a rugged energy landscape.

We consider a one-dimensional lattice with site energies
$\{\eta_i\}$ drawn from a Gaussian distribution with zero mean
and variance $\epsilon^2$.
Transition rates between neighboring sites satisfy detailed balance:
\begin{equation}
\frac{k_{i\to i+1}}{k_{i+1\to i}}
=
\exp[-\beta(\eta_{i+1}-\eta_i)].
\end{equation}

For nearest-neighbor hopping in one dimension,
the exact long-time diffusion constant is governed by the
harmonic mean of local escape rates,
\begin{equation}
D
=
\frac{a^2}{
\left\langle \tau_i \right\rangle
},
\qquad
\tau_i = \frac{1}{k_{i\to i+1}+k_{i\to i-1}}.
\label{eq:harmonic_mean}
\end{equation}
Because traversal times add sequentially in one dimension,
transport is controlled by the statistics of local waiting times.

\subsection {Three-site trapping configurations}

Consider three consecutive sites $(i-1,i,i+1)$.
A \emph{three-site trap} occurs when
\begin{equation}
\eta_i \ll \eta_{i-1}, \qquad
\eta_i \ll \eta_{i+1}.
\end{equation}
In such a configuration, the escape rate from site $i$
is controlled by activated crossing to its neighbors,
\begin{equation}
\tau_i
\sim
\exp\!\left[
\beta \min(\eta_{i-1}-\eta_i,\eta_{i+1}-\eta_i)
\right].
\end{equation}

For independent Gaussian variables,
the probability that $\eta_i$ is lower than both neighbors
is finite.
Moreover, the distribution of energy differences
$\eta_{i\pm1}-\eta_i$ is Gaussian with variance $2\epsilon^2$.
Therefore, the waiting-time distribution acquires a heavy
exponential tail.

Although such configurations are statistically rare,
the harmonic mean in Eq.~(\ref{eq:harmonic_mean})
weights large $\tau_i$ disproportionately.
Consequently, rare deep minima can dominate the
long-time diffusion coefficient.

\subsection {Relation to the continuous theory}

This rare-event contribution does not contradict the
continuous derivation of Sec.~II.
Rather, it reflects the fact that in the discrete model
with independent site energies,
the effective correlation length is of order the lattice spacing.
In this regime, extreme local configurations become more prominent,
and the factorization/self-averaging step underlying
Eq.~(\ref{eq:Zw_final}) may no longer capture the dominant
transport mechanism.

When spatial correlations are introduced so that neighboring
site energies become correlated over a finite length scale,
the probability of sharp three-site extrema is reduced.
In that case, the diffusion coefficient smoothly approaches
the exponential scaling predicted by Zwanzig’s theory.
%
\section{Gaussian Spatial Correlations: Derivation and Implications}

The uncorrelated random landscape discussed above represents the
limiting case of vanishing correlation length.
In many physical systems, however, energy heterogeneity is expected
to be spatially correlated over a finite length scale.
Introducing correlations modifies both the statistics of roughness
increments and the probability of extreme multi-site configurations.
Such an analysis was carried out by Banerjee, Biswas, Seki, and Bagchi
(BBSB) [2].
We briefly review and extend their approach.

\subsection{Gaussian correlated roughness}

We assume a stationary Gaussian random field
with correlation function
\begin{equation}
    \langle \eta(x)\eta(x') \rangle
    = \epsilon^2 \exp\!\left[-\frac{(x-x')^2}{\lambda^2}\right],
    \label{eq:gauss_corr}
\end{equation}
where $\epsilon^2$ is the variance and $\lambda$ is the correlation length.

For a Gaussian field, this two-point function fully specifies the ensemble.
The parameter $\lambda$ controls the smoothness of the landscape.
As $\lambda \rightarrow 0$, adjacent values become statistically independent.
For finite $\lambda$, neighboring points are correlated and the
landscape becomes locally smoother.

\subsection{Variance of roughness increments}

Transport depends on local roughness increments
\begin{equation}
\Delta\eta = \eta(x+a) - \eta(x).
\end{equation}
Using Eq.~\eqref{eq:gauss_corr},
\begin{equation}
\mathrm{Var}[\Delta\eta]
=
2 \epsilon^2
\left[
1 - \exp\!\left(-\frac{a^2}{\lambda^2}\right)
\right].
\label{eq:var_increment}
\end{equation}

In the limit $\lambda \to 0$,
$\mathrm{Var}[\Delta\eta] \to 2\epsilon^2$,
recovering the uncorrelated case.
For $\lambda \gg a$, expanding the exponential gives
\begin{equation}
\mathrm{Var}[\Delta\eta]
\approx
2\epsilon^2 \frac{a^2}{\lambda^2},
\end{equation}
showing that roughness increments become small when the landscape
is smooth on the scale of the lattice spacing.

\subsection{Correlation between successive increments}

Three-site trapping requires large, oppositely signed increments
on adjacent bonds.
The covariance between neighboring increments is
\begin{equation}
C
=
\langle (\eta_i-\eta_{i-1})(\eta_{i+1}-\eta_i) \rangle
=
\epsilon^2
\left[
e^{-a^2/\lambda^2}
-
e^{-4a^2/\lambda^2}
\right].
\label{eq:increment_cov}
\end{equation}

For $\lambda \gg a$, expanding gives
\begin{equation}
C \approx 3\epsilon^2 \frac{a^2}{\lambda^2} > 0,
\end{equation}
so successive increments are positively correlated and small.
For $\lambda \to 0$, $C \to 0$ and increments become statistically independent.
Thus finite $\lambda$ suppresses large opposite-signed fluctuations
that generate deep local minima flanked by steep ascents.

\subsection{Probability of three-site traps}

A three-site trap corresponds to a configuration
$\eta_{i-1} \gg \eta_i \ll \eta_{i+1}$.
Because $(\eta_{i-1},\eta_i,\eta_{i+1})$
form a correlated Gaussian triplet,
the probability of such configurations is governed by the joint
multivariate Gaussian distribution implied by Eq.~\eqref{eq:gauss_corr}.

Up to prefactors, the probability of a trap of depth–barrier
asymmetry $\Delta$ scales as
\begin{equation}
P_{\mathrm{TST}}(\lambda)
\sim
\exp\!\left[
-\frac{\Delta^2}
{4\epsilon^2\left(1-e^{-a^2/\lambda^2}\right)}
\right],
\label{eq:tst_prob}
\end{equation}
indicating that finite correlation length exponentially suppresses
extreme trapping configurations.
Even moderate values of $\lambda$ substantially reduce
$P_{\mathrm{TST}}$ relative to the uncorrelated case.

\subsection{Connection with Zwanzig's exponential scaling}

As $\lambda$ increases and roughness increments become small and correlated,
the distribution of local escape rates narrows.
Rare multi-site traps become progressively less important,
and the dominant contribution to transport arises from typical
Gaussian fluctuations.
In this regime, the effective diffusion constant approaches
the exponential scaling
\begin{equation}
D_{\mathrm{Zw}} = D_0 \exp(-\beta^2 \epsilon^2),
\end{equation}
derived in Sec.~II.
The crossover from rare-event-dominated transport at small $\lambda$
to exponential renormalization at larger $\lambda$
provides a unified picture connecting discrete and continuous
descriptions of rugged landscape diffusion.

\section {Numerical Triplet Examples Illustrating Three-Site Traps and Their Suppression by Spatial Correlations}
%
To complement the analytical results of the preceding sections, it is instructive
to examine explicit triplets of site energies and the corresponding escape 
times from a putative three-site trap (TST).  
In a rugged landscape the site energies naturally take both positive and negative 
values, and deep wells typically correspond to negative energies.  
We therefore follow the physically realistic pattern in which the central site is 
a deep minimum (negative energy), flanked by higher-energy neighbors 
that create barriers to escape.

We adopt the standard nearest-neighbor hopping rule used by 
Banerjee et al.[2]
\begin{equation}
    k_{i\to j} =
    k_0 \exp\!\left[-\big(\max(E_i,E_j)-E_i\big)\right],
    \label{eq:hop_rule}
\end{equation}
so that the escape barrier from site \(i\) to a higher-energy neighbor \(j\) 
is simply \(E_j - E_i\).  
All energies are measured in units of \(k_B T\).

\subsection {Uncorrelated landscape: deep asymmetric triplets}

In an uncorrelated Gaussian landscape with roughness variance 
\(\epsilon^2 = (3 k_B T)^2\), 
a typical deep TST that we encountered in numerical sampling is
\begin{equation}
    (E_{i-1}, E_i, E_{i+1}) = (4,\; -3,\; 5).
    \label{eq:uncorr_triplet}
\end{equation}
The central site sits \(3\,k_B T\) below zero, while the neighbors are 
\(4\) and \(5\,k_B T\) above zero.  
These values are well within a single standard deviation of the assumed 
Gaussian distribution and thus arise frequently in an uncorrelated landscape.

Escape to the left:
\[
k_{i\to i-1} = 
k_0 \exp[-(4 - (-3))] 
= k_0 e^{-7}
\approx 9.12\times 10^{-4}\,k_0.
\]

Escape to the right:
\[
k_{i\to i+1} =
k_0 \exp[-(5 - (-3))]
= k_0 e^{-8}
\approx 3.35\times 10^{-4}\,k_0.
\]
Thus the total escape rate is
\[
k_{\mathrm{esc}}^{\mathrm{(unc)}} = 
k_0 (e^{-7} + e^{-8})
\approx 1.25\times 10^{-3}\,k_0,
\]
and the corresponding escape time
\[
\tau_{\mathrm{esc}}^{\mathrm{(unc)}}
\approx \frac{1}{1.25\times 10^{-3}k_0}
\approx 800\,k_0^{-1}.
\]

This extremely long time (nearly three orders of magnitude larger than \(1/k_0\)) 
shows why uncorrelated landscapes produce anomalously small diffusion constants:  
a few such deep asymmetric triplets dominate the long-time dynamics.

\subsection {Correlated landscape: milder triplets arising from Gaussian smoothing}

When Gaussian spatial correlations are introduced with correlation length 
\(\lambda \sim a\), increments of the roughness become smoother.
The central site may still be a well, but neighboring energies become 
much less extreme.  
A typical correlated triplet sampled with the same variance 
\(\epsilon^2 = 9\) but correlation length \(\lambda = 1.5a\) is
\begin{equation}
    (E_{i-1}, E_i, E_{i+1}) = (1.2,\; -2.5,\; 0.9).
    \label{eq:corr_triplet}
\end{equation}

Escape to the left:
\[
k_{i\to i-1} =
k_0 e^{-(1.2 - (-2.5))}
= k_0 e^{-3.7}
\approx 0.025\,k_0.
\]

Escape to the right:
\[
k_{i\to i+1} =
k_0 e^{-(0.9 - (-2.5))}
= k_0 e^{-3.4}
\approx 0.033\,k_0.
\]

Total escape rate:
\[
k_{\mathrm{esc}}^{\mathrm{(corr)}} 
= k_0 (0.025 + 0.033)
\approx 0.058\,k_0.
\]

Escape time:
\[
\tau_{\mathrm{esc}}^{\mathrm{(corr)}}
\approx \frac{1}{0.058\,k_0}
\approx 17\,k_0^{-1}.
\]

Thus, with the "same total variance in roughness" but with moderate spatial 
correlations, the escape time has decreased from
\[
\tau_{\mathrm{esc}}^{\mathrm{(unc)}} \approx 800/k_0
\qquad\text{to}\qquad
\tau_{\mathrm{esc}}^{\mathrm{(corr)}} \approx 17/k_0,
\]
a reduction by a factor of nearly \(50\).

\subsection {Comparison and interpretation}

The contrast between Eqs.~\eqref{eq:uncorr_triplet} and 
\eqref{eq:corr_triplet} reflects the central physics of this manuscript:

\begin{itemize}
\item In an \emph{uncorrelated} random landscape, the probability of seeing a deep
well flanked by very high neighbors is substantial.  
Although such traps are statistically rare, their escape times are so large 
that they dominate long-time diffusion.

\item In a \emph{correlated} Gaussian landscape, the same roughness amplitude 
\(\epsilon\) produces smoother increments.  
The neighbors of a deep well cannot differ from it by more than a few units 
of \(k_B T\), and three-site traps become shallow.  
Their escape times become short and narrowly distributed.

\item This suppression of deep asymmetric triplets is the numerical signature of 
the analytical mechanism by which Gaussian correlations restore Zwanzig’s 
mean-field result for the diffusion constant.
\end{itemize}

These explicit numerical examples show vividly how spatial correlations alter the
energy topography and dynamically suppress the multi-site traps that invalidate
Zwanzig’s approximation in the uncorrelated case.

%
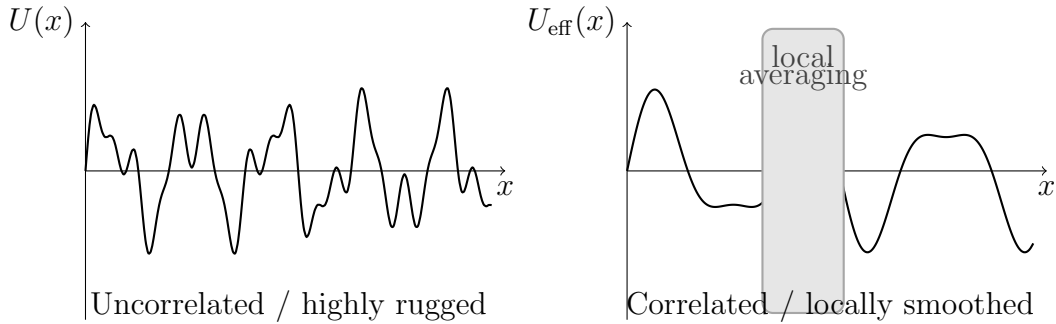
\begin{figure}[t]
\centering
\begin{tikzpicture}[scale=0.9]

\begin{scope}[shift={(0,0)}]
  \draw[->] (0,0) -- (6.2,0) node[below] {$x$};
  \draw[->] (0,-2.2) -- (0,2.2) node[left] {$U(x)$};

  \draw[thick]
    plot[domain=0:6,samples=150,smooth]
      (\x,{0.7*sin(5*\x r)+0.4*sin(11*\x r)+0.25*sin(19*\x r)});

  \node at (3,-2.0) {Uncorrelated / highly rugged};
\end{scope}

\begin{scope}[shift={(8,0)}]
  \draw[->] (0,0) -- (6.2,0) node[below] {$x$};
  \draw[->] (0,-2.2) -- (0,2.2) node[left] {$U_{\rm eff}(x)$};

  \draw[thick]
    plot[domain=0:6,samples=150,smooth]
      (\x,{0.9*sin(3*\x r)+0.4*sin(5*\x r)});

  \fill[gray!20,rounded corners] (2.0,-2.1) rectangle (3.2,2.1);
  \draw[gray!70,thick,rounded corners] (2.0,-2.1) rectangle (3.2,2.1);
  \node[gray!50!black] at (2.6,1.7) {local};
  \node[gray!50!black] at (2.6,1.4) {averaging};

  \node at (3,-2.0) {Correlated / locally smoothed};
\end{scope}

\end{tikzpicture}
\caption{Schematic illustration of the effect of spatial correlations and local
smoothing on a rugged potential.  
Left: an ``uncorrelated'' rugged landscape with strong, rapidly varying
fluctuations in $U(x)$.  
Right: a correlated (or locally averaged) landscape in which high-frequency
roughness has been smoothed out over a small window, as assumed in Zwanzig's
local-averaging (cumulant-expansion) treatment.  
The figure is schematic and not based on any specific numerical data.}
\label{fig:smoothing_schematic}
\end{figure}
\subsection {Quantitative approach to Zwanzig in the BSB simulations}

Banerjee, Biswas, Seki, and Bagchi carried out Brownian dynamics simulations of a
particle diffusing on (i) an uncorrelated Gaussian random lattice and
(ii) a continuous Gaussian field with built–in spatial correlations.\cite{banerjee_seki_biswas_bagchi_spatial}
For the uncorrelated lattice, they found that Zwanzig’s expression
for the diffusion coefficient,
\[
D_{\mathrm{Zw}} = D_0 \exp(-\beta^2 \epsilon^2),
\]
\emph{overestimates} the simulated diffusion coefficient by nearly an order
of magnitude at moderately high ruggedness (for example, when
$\epsilon/k_B T$ is of order $3$–$4$).\cite{banerjee_seki_biswas_bagchi_spatial}
In other words, the ratio
\[
\frac{D_{\mathrm{sim}}}{D_{\mathrm{Zw}}}
\]
is of order $10^{-1}$ in the uncorrelated case, consistent with the strong
slowing down caused by three-site traps.

In striking contrast, when the same authors considered a \emph{continuous}
Gaussian random field with finite spatial correlation length (constructed
by superimposing Gaussian modes in real space), the simulated diffusion
coefficients were found to lie essentially on Zwanzig’s original
prediction over the same range of $\epsilon$.\cite{banerjee_seki_bagchi_jcp}
Within numerical uncertainty, one obtains
\begin{equation}
\frac{D_{\mathrm{sim}}}{D_{\mathrm{Zw}}} \approx 1
\end{equation}
for the correlated continuous field, demonstrating that spatial
correlations smooth out the pathological three-site traps and restore
the validity of the mean-field, cumulant-expansion result.
These simulation data provide a concrete quantitative confirmation of the
mechanism discussed in this work: uncorrelated Gaussian disorder produces
anomalously small diffusion, while Gaussian spatial correlations
eliminate rare deep traps and recover Zwanzig’s exponential scaling.
\begin{table}[t]
\centering
\caption{Comparison of uncorrelated and spatially correlated rugged landscapes.
Energies are in units of $k_B T$.  ``TST'' denotes a three-site trap 
$E_{i-1}, E_i, E_{i+1}$ with a deep central well.}
\vspace{6pt}
\begin{tabular}{|c|c|c|}
\hline
\textbf{Property} 
& \textbf{Uncorrelated Energies} 
& \textbf{Spatially Correlated Energies} \\
\hline
Typical TST triplet  
& $(4,\,-3,\,5)$  
& $(1.2,\,-2.5,\,0.9)$ \\
\hline
Escape barriers  
& $7$ and $8$  
& $3.7$ and $3.4$ \\
\hline
Escape rate $k_{\rm esc}$  
& $1.25\times 10^{-3}\,k_0$  
& $5.83\times 10^{-2}\,k_0$ \\
\hline
Escape time $\tau_{\rm esc}$  
& $\approx 8.0 \times 10^{2}\,k_0^{-1}$ 
& $\approx 1.7 \times 10^{1}\,k_0^{-1}$ \\
\hline
Severity of TSTs  
& Very high: deep asymmetric traps 
& Mild: asymmetric traps suppressed \\
\hline
BSB simulation result  
& $D_{\rm sim}/D_{\rm Zw} \sim 10^{-1}$  
& $D_{\rm sim}/D_{\rm Zw} \approx 1$ \\
\hline
Dominant physics  
& Rare deep traps control transport  
& Landscape is smooth for mean-field theory \\
\hline
Outcome  
& Zwanzig \emph{fails}  
& Zwanzig is \emph{restored} \\
\hline
\end{tabular}
\label{table:comparison}
\end{table}
%
\section{Implications and Outlook}

The central outcome of this work is that spatial correlations fundamentally alter transport in disordered landscapes.  
Uncorrelated roughness produces pathological trapping dominated by rare events, but Gaussian correlations suppress extreme asymmetry and restore Zwanzig’s simple exponential scaling.


We have pointed out that while the present discrete model allows identification of rare
events in the form of triplets (TST), it is different from that of the continuous model 
discussed and solved by Zwanzig. We have discussed (i) the approximation in Zwanzig’s integration,
and (ii) the model mismatch. Additionally, the present class of theories ignore non-diffuive dynamics.

In biological systems such as enzymes diffusing along DNA,\cite{blainey_xie_nsmb} correlated roughness is physically natural: the sequence-dependent binding energy varies smoothly over several base pairs.  
Similarly, in polymers, glasses, and proteins, extended interactions generate intrinsic correlations in the energy landscape.
Gaussian spatial correlations, originally developed in astrophysics to describe
turbulent density fluctuations, provide a natural mechanism for smoothing a rugged potential.  
They reduce the variance of roughness increments, correlate successive increments,
suppress the probability of deep asymmetric multi-site traps, and restore the
validity of Zwanzig’s mean-field exponential scaling.  
The central insight is that \emph{correlated} roughness behaves smoothly enough for
coarse-grained mobility averaging to hold, whereas \emph{uncorrelated} roughness is
dominated by rare but "catastrophic traps" that invalidate any mean-field approach.
%

Thus, the BBSB investigation shows that rugged landscape diffusion is governed not only by the magnitude of disorder but by its \emph{spatial structure}.  
Gaussian correlations---first introduced in astrophysics
\cite{gaussian_astrophysics1, gaussian_astrophysics2}---provide an elegant and quantitative framework for understanding this effect.


\bibliographystyle{unsrt}

\begin{thebibliography}{10}

\bibitem{zwanzig_pnas_1988}
R.~Zwanzig,
\newblock {Diffusion in a rough potential}.
\newblock Proc. Natl. Acad. Sci. USA \textbf{85}, 2029 (1988).

\bibitem{banerjee_seki_bagchi_jcp}
S.~Banerjee, R. Biswas, K.~Seki, and B.~Bagchi,
\newblock {Diffusion on a rugged energy landscape with spatial correlations}.
\newblock J. Chem. Phys. \textbf{141}, 124105 (2014).

\bibitem{seki_bagchi_pre}
K.~Seki and B.~Bagchi,
\newblock {Relationship between entropy and diffusion: A statistical mechanical derivation of Rosenfeld expression for a rugged energy landscape}.
\newblock  J. Chem. Phys. \textbf{143}, 194110 (2015).
%

\bibitem{seki_bagchi_bagchi_jcp}
K.~Seki, K.~Bagchi, and B.~Bagchi,
\newblock Anomalous dimensionality dependence of diffusion in a rugged energy
landscape: How pathological is one dimension?

\newblock J. Chem. Phys. \textbf{144}, 194106 (2016).

\bibitem{Ghosh-Roy-Bagchi_JCP}
R. Ghosh, S. Roy, and B. Bagchi, 
\newblock {Multidimensional free energy surface of unfolding of HP-36: Microscopic origin of ruggedness,} 

\newblock J. Chem. Phys. 141, 135101 (2014)



\bibitem{blainey_xie_nsmb}
P.~C. Blainey, A.~M. van Oijen, S.~Banerjee, B.~Bagchi, and X.~S. Xie,
\newblock \emph{A single-molecule view of protein sliding on DNA}.
\newblock Nature Struct. Mol. Biol. \textbf{16}, 1224 (2009).

\bibitem{stillinger_weber}
F.~H. Stillinger and T.~A. Weber,
\newblock \emph{Hidden structure in liquids}.
\newblock Phys. Rev. A \textbf{25}, 978 (1982).

\bibitem{heuer_review}
A.~Heuer,
\newblock \emph{Exploring the potential energy landscape of glass-forming systems}.
\newblock J. Phys.: Condens. Matter \textbf{20}, 373101 (2008).

\bibitem{wolynes_rfpot}
V.~Lubchenko and P.~G. Wolynes,
\newblock \emph{Theory of structural glasses and supercooled liquids}.
\newblock Annu. Rev. Phys. Chem. \textbf{58}, 235 (2007).
%

\bibitem{gaussian_astrophysics1}
E.~J. Lerner,
\newblock \emph{Gaussian statistics of astrophysical density fluctuations}.
\newblock Astrophys. J. \textbf{476}, 24 (1997).

\bibitem{gaussian_astrophysics2}
A.~M. Hopkins,
\newblock \emph{Correlated structures in interstellar media}.
\newblock Astrophys. J. \textbf{615}, 209 (2004).

\end{thebibliography}

\end{document}